\documentclass[twoside,twocolumn,a4paper]{IEEEtran}

\usepackage{amsfonts}
\usepackage{amsmath}
\usepackage{mathtools}
\usepackage{bm,upgreek}      
\usepackage{color}
\usepackage{url}

\usepackage{setspace}
\usepackage{helvet}         
\usepackage{courier}       
\usepackage{type1cm}       

\usepackage[bottom]{footmisc}
\usepackage{multicol}

\usepackage{color}

 \def\gmat#1{\mbox{\boldmath$#1$}}
\newcommand \dd[1]  { \,\textrm d{#1}                       }

\newcommand{\be}{\begin{equation}}
\newcommand{\ee}{\end{equation}}

\begin{document}

\title{Perspective: Quantum Hamiltonians for Optical Interactions}
\author {David L. Andrews\thanks{School of Chemistry, University of East Anglia, Norwich Research Park, Norwich NR4 7TJ, U.K.{email:d.l.andrews@uea.ac.uk}},~~Garth A. Jones\thanks{School of Chemistry, University of East Anglia, Norwich Research Park, Norwich NR4 7TJ, U.K. {email:garth.jones@uea.ac.uk}},~~A Salam\thanks{Department of Chemistry, Wake Forest University, Winston-Salem, NC 27109, USA.{email:salama@wfu.edu}},~~and R. Guy Woolley\thanks{School of Science and Technology, Nottingham Trent University, Nottingham NG11 8NS, U.K.{email: rguywoolley@cantab.net}}
\thanks{Accepted for publication in Journal of Chemical Physics, January 2018, as a \textit{Perspective}.}}

\maketitle

\begin{abstract}
The multipolar Hamiltonian of quantum electrodynamics (QED) is extensively employed in chemical and optical physics to treat rigorously the interaction of electromagnetic fields with matter.  It is also widely used to evaluate intermolecular interactions.  The multipolar version of the Hamiltonian is commonly obtained by carrying out a unitary transformation of the Coulomb gauge Hamiltonian that goes by the name of Power-Zienau-Woolley (PZW). Not only does the formulation provide excellent agreement with experiment, and versatility in its predictive ability, but also superior physical insight. Recently, the foundations and validity of the PZW Hamiltonian have been questioned, raising a concern over issues of gauge transformation and invariance, and whether observable quantities obtained from unitarily equivalent Hamiltonians are identical. Here, an in-depth analysis of theoretical foundations clarifies the issues and enables misconceptions to be identified.  Claims of non-physicality are refuted: the PZW transformation and ensuing Hamiltonian are shown to rest on solid physical principles and secure theoretical ground.
\end{abstract}
\pagestyle{plain}
\section{Introduction}
\label{Intro}
In the rapidly developing and expanding field of modern photonics, it is appropriate to revisit periodically key elements of the underlying theory. Challenging received wisdom and re-evaluating its validity can reveal fresh insights and potentially expose flaws. When it is necessary to account  correctly for the interaction of electromagnetic radiation with atoms and molecules on the quantum scale, one advantageous formulation of fundamental theory is the casting of the Hamiltonian in terms of the purely transverse electromagnetic field variables ${\bf E}^{\perp}, {\bf B}$ associated with radiation. This is possible both in classical electrodynamics in Hamiltonian form and in quantum electrodynamics (QED). The quantum case has long been known in the literature as the Power-Zienau-Woolley (PZW) Hamiltonian \cite{PZ:59,AW:70,RGW:71,EAP:64,WPH:82,RL:83,TDG:89,SM:95,CT:98,AS:10}. It arises from a particular transformation of the familiar Coulomb gauge Hamiltonian, generated by the unitary operator
\begin{equation}
    \mathsf{U} = \exp(i\mathsf{S}/\hbar)
    \label{1.1U},
\end{equation}
where
\begin{equation}
\mathsf{S}~=~\int\!{\bm{\mathsf{P}}}\cdot{\bm{\mathsf{A}}}~\dd\tau.
\label{action}
\end{equation}
Here, $\mathsf{S}$ signifies a coupling of electromagnetic and material variables: ${\bm{\mathsf{A}}}$ 
is the Coulomb gauge vector potential operator, and ${\bm{\mathsf{P}}}$ is an operator solution of the equation
\begin{equation}
    \nabla\cdot{\bm{\mathsf{P}}}~=~-\uprho,
    \label{Prho}
\end{equation}
where $\uprho$ is the molecular charge density operator. The multipolar Hamiltonian is the special case obtained when ${\bm{\mathsf{P}}}$ is represented as an expansion in multipole moment operators\cite{AW:70}; in practical applications it is usually only the leading terms of the multipole series that need be retained.

In this \textit{Perspective} we address several misconceptions recently raised in the literature 
regarding this multipolar form of the QED Hamiltonian, widely used in atomic, 
molecular and optical physics, and theoretical chemistry. 
These issues have resurfaced as the focus of a recent paper by Rousseau and 
Felbecq \cite{RF:17} (hereafter referred to as RF). Their article aimed to show that a 
quantum formulation of electrodynamics that is based on the PZW Hamiltonian is fundamentally 
flawed and therefore unsuited to its many applications. This is a surprising claim, since over 
its close to sixty year history there has been a host of applications 
to which the PZW Hamiltonian has been applied, which have agreed with, or led to predictions borne 
out by experiment. There are many hundreds of papers in the established literature citing the original work, bearing testimony to its efficacy and success; numerous significant applications and advances have built upon it, even over the last decade \cite{JJ:17,NQ:17,AD:16,RG:16,AS:16a,GM:15,VYC:15,MDT:15,KAF:15,AS:15,SP:14,YK:14,AS:13,ACJ:13,ENA:13,AS:12,RM:10,DLA:10,MK:10,BSS:09,RV:09,SH:08,SARH:08,KKC:08,PGB:08,BS:08}. Indeed we know of no cases where the theory has been faulted by experimental study - which would usually be the condition to 
invite reappraisal of a previously successful theory.  

Specifically, RF claim that the Coulomb gauge and multipolar Hamiltonians ``predict different physical results'', even though they have given no calculation of a physical observable to show this. It may be noted at once that the two Hamiltonians are related by the transformation (\ref{1.1U}), and that all the standard results about unitary transformations in quantum mechanics apply; furthermore, the generator $\mathsf{S}$ obviously commutes with the vector potential operator ${\bm{\mathsf{A}}}$ and the Coulomb gauge condition, $\nabla\cdot{\bm{\mathsf{A}}} = 0$ which appears as a constraint in the Hamiltonian theory. These remarks will be elaborated upon in \S\ref{genHam} and \S \ref{PZW}.

We also note that the applicability of a PZW-like transformation has been demonstrated in high-energy physics (in non-abelian gauge theory)\cite{RQB:12,HJ:97}, and in the proof of the stability of ordinary matter in the presence of quantized radiation - i.e. in the real world, through rigorous functional analysis\cite{LL:03}. In these cases the intention is to be able to pass between the Coulomb and PZW formulations as convenient; for example, the PZW transformed version has smoother mathematical behaviour at low photon frequencies - since it involves the electromagnetic fields rather than the vector potential - as compared to the Coulomb gauge form, and this facilitates certain important estimates in the proof of stability. It has been used in its multipolar form to describe QED in a cavity\cite{PT:82} - see also\cite{VGD:14,VGD:15}. It has also been applied to the Coulomb gauge Hamiltonian describing a medium that is linear, absorbing and dispersing to generate a multipolar formulation of macroscopic QED, successfully employed in the calculation of Casimir-van der Waals forces\cite{SYB:04,SYB:12}. 

In this work we focus on the challenge, demonstrating that any assertion that ``the PZW Hamiltonian has inconsistencies'' is simply wrong; we identify serious inaccuracies in the analysis offered in support of this recent claim\cite{RF:17}.  Furthermore, we clarify several points regarding important physical notions relating to near-field, multipolar quantum electrodynamics, that may have led to these erroneous conclusions. Our examination reaffirms the rigour, internal consistency and validity of the PZW transformation and the ensuing Hamiltonian. 
At a time when the span of application for photonic interactions is developing apace, we believe it is crucial for theorists, computational modellers and experimentalists alike to have confidence in the rigour and correctness of core theory.  

One of the less recognized problems when working within the framework of \textit{semiclassical} radiation theory is the temptation to cultivate a view of `gauge transformations' exactly as in Maxwell's theory, failing to appreciate that through the construction of a quantum Hamiltonian from a classical Lagrangian, using the general method due originally to Dirac\cite{PAMD:52}: 1) \textit{the scalar potential is dispensed with completely}, and 2) \textit{there are no time-dependent gauge transformations}.  Temporal evolution depends on a vector potential with no explicit time-dependence, and its conjugate momentum, simply through modified Poisson-brackets (known as Dirac brackets) which become commutators after quantization. This will be described in \S \ref{genHam}, which presents the fundamental theory related to quantum optical Hamiltonians. Then in \S \ref{PZW} we describe some specifics of the PZW transformation. Once this has been developed, we will be in a position to clarify misconceptions about the applicability of the PZW Hamiltonian, as well as certain subtleties when working in near-field QED applications: this will be carried out in \S \ref{Discuss}. Where we can, we will also attempt to give insight into how we think some of these common misunderstandings may have arisen. Finally, in \S \ref{out} we give some indications of the present scope and future prospects for application of the multipolar Hamiltonian.

\section{The general Hamiltonian}
\label{genHam}
The cornerstone of modern theories of the interactions of atoms and molecules 
with electromagnetic radiation is a classical Lagrangian that has been known 
for more than a hundred years. For particles with charge and mass parameters 
\{$e_{n}, m_{n}, ~ n= 1, \ldots N$\} and positions \{${\bf x}_{n}, n = 1, \ldots
N$\} interacting with an electromagnetic field (${\bf E}, {\bf B}$) 
this is\cite{KS:03, WH:36, EAP:64}
\begin{align}
  L&=\tfrac{1}{2}\sum_n m_n |\dot{\bf x}_n|^{2}-\int\rho\phi \dd \tau+\int{\bf j}\cdot{\bf a} \dd \tau\nonumber \\
&+ \tfrac{1}{2}\epsilon_0\int\big({\bf
E}\cdot{\bf E}-c^2{\bf B}\cdot{\bf B}\big)\dd \tau,
\label{gH.1}
 \end{align}
where $\phi,{\bf a}$ are field potentials satisfying
\begin{equation}
 {\bf E} = -\nabla \phi - \frac{\partial {\bf a}}{\partial t}
 \label{gH.2},~~~~{\bf B}= \nabla \wedge {\bf a},
\end{equation}
but not otherwise restricted (that is, no gauge condition is implied). The variables 
$\rho$ and ${\bf j}$ are the usual charge and current densities for point particles, defined as the distributions
\begin {equation}
 \rho({\bf x})= \sum_{n}e_{n}\delta^{3}({\bf x}_{n}-{\bf x}),~~~
{\bf j}({\bf x})=\sum_{n}e_{n}\dot{\bf x}_{n}\delta^{3}({\bf x}_{n}-{\bf x}).
 \label{gH.3}
\end{equation}

Taking \{${\bf x}_{n},\dot{\bf x}_{n}$, \textit{n}=1\ldots \textit{N}\} and the field potentials, $\phi$ and ${\bf a}$,
as the independent dynamical variables in the Lagrangian function for the interaction of charged particles with a radiation field, and substituting $L$ in the Euler-Lagrange equations yields the Maxwell equations and Lorentz Force Law as the equations 
of motion. This Lagrangian describes a closed system and 
$\partial L/\partial t =0$; hence the classical Hamiltonian $H$ derived from it by the usual arguments is 
the constant energy of the whole system. Any Lagrangian that does not contain \textit{derivatives} 
of the field strengths $({\bf E}, {\bf B})$ is said to show 
\textit{minimal coupling} between the charges and the fields, independently 
of any gauge condition that might be invoked. While it is possible to 
develop a quantum theory in this Lagrangian framework using Feynman's 
path integral formalism\cite{HN:95}, it is customary in non-relativistic QED to work with the Schr\"{o}dinger equation for a Hamiltonian operator. Usually this Hamiltonian is obtained by following the
canonical quantization prescription, due to Dirac, of first obtaining a classical Hamiltonian theory, and then reinterpreting the Poisson-bracket algebra of the variables as quantum commutators.  We follow this traditional route.

There is an immediate technical difficulty, however, for the formulation of electrodynamics 
in an appropriate Hamiltonian framework starting from (\ref{gH.1}). The canonical momenta for the electromagnetic 
field described by arbitrary potentials $\phi,{\bf a}$ consistent 
with (\ref{gH.2}), are defined as 
$\gmat{\pi} = \delta L/\delta \dot{\bf a}$, $\pi_{0}=\delta L/\delta 
\dot{\phi}$. Since $L$ has no terms in $\dot{\phi},~\pi_{0}=0$; this implies 
the Legendre transformation from Lagrangian to Hamiltonian variables is 
singular. The traditional solution to this problem, due to Fermi\cite{EF:32}, involved the modification of
$L$ by the introduction of a gauge condition which does not affect the equations of motion. 
This is now viewed as a historical procedure 
and modern Hamiltonian theory\cite{SW:95,RR:10,SW:13} derives from the general solution to this problem
first proposed by Dirac\cite{PAMD:50,PAMD:51,PAMD:52,PAMD:64}; it involves the elimination of the scalar 
potential from the formalism and the recognition of the existence of `constraints' (relations between the canonical variables). The application of his
method to the electrodynamics of atoms and molecules provided for subsequent development in the 
chemical physics literature\cite{AW:70, RGW:71, RGW:75a,
RGW:75, RGW:03}. In Dirac's original work the Hamiltonian is 
built out of only manifestly \textit{gauge-invariant} quantities\cite{PAMD:55}.

In the non-relativistic electrodynamics of charged particles, the 
gauge-invariant classical Hamiltonian that follows from Dirac's method applied to 
$L$, equation (\ref{gH.1}), is\cite{RGW:03}
\begin{equation}
 H~=~\sum_{n}\frac{1}{2m_n}|{\overline{\bf p}}_{n}|^{2}~+~\tfrac{1}{2}\epsilon_0\int\!\!\bigg(|{\bf E}|^2~+~c^2|{\bf B}|^2\bigg)\dd \tau
 \label{gH.4}
\end{equation}
where, for each of the particles labeled \textit{n}, $\overline{\bf p}_{n}= {\bf p}_{n} - e_{n}{\bf a}({\bf x}_{n})$, and $\epsilon_{0}{\bf E} = -{\gmat{\pi}}$ in terms of the canonical variables \{${\bf x}_{n},{\bf p}_{n},{\bf a}, {\gmat{\pi}}\}$. 
Superficially this appears to be the Hamiltonian for `free' charges and the electromagnetic field. However, the charge-field coupling parameter $e_n$ appears in 
the non-zero Poisson brackets
of the \{$\overline{\bf p}_{n}$\} variables which are not canonical variables,
\begin{align}
 \{{\overline{\rm p}}_{n}^{\alpha},{\overline{\rm p}}_{n}^{\beta}\}&=e{_n}{\epsilon}_{\beta\alpha\mu}B({\bf x}_{n})^{\mu},\nonumber \\
\{{\overline{\rm p}}_{n}^{\alpha}, E({\bf x})^{\mu}\}&=e_{n}\epsilon_{0}^{-1}\delta_{\alpha\mu}\delta^3({\bf x}_{n}-{\bf x}).
 \label{gH.5}
\end{align}
To these must be conjoined the conditions
\begin{align}
\{x_{n}^{\alpha}, \overline{\rm p}_{m}^{\beta}\} = ~& \delta_{nm}\delta_{\alpha\beta}, \nonumber\\
\{B({\bf x})^{\beta},E({\bf x}')^{\mu}\}=&-\epsilon_{0}^{-1}\epsilon_{\beta\alpha\mu}\nabla_{\bf x}^{\alpha}\delta^3({\bf x}-{\bf x}'),
\label{gH.6}
\end{align}
where $\epsilon_{\alpha\beta\mu}$ is the usual Levi-Civita symbol and the Greek letters label the components of the 3-dimensional vectors.   
The inter-charge Coulomb interaction becomes explicit when the electric field is split into longitudinal (${\bf E}^{\parallel}$) and transverse (${\bf E}^{\perp}$) components - a gauge-invariant separation. The space integral of $|{\bf E}^{\parallel}|^2$ gives the Coulomb energies, and the integral over $|{\bf E}^{\perp}|^2$ is associated with radiation. Equations (\ref{gH.4})-(\ref{gH.6}) constitute a complete statement of the 
classical electrodynamics of charged particles in Hamiltonian form.

The appropriate representation of atoms and photons becomes possible only after quantization of the 
particle and field variables. Canonical
quantization postulates a direct correspondence between classical
Poisson brackets and quantum commutators, applied to all observables
\begin{equation}
  \{A,B\}=C ~ \Rightarrow  ~[\mathsf{A}, \mathsf{B}]=i\hbar ~\mathsf{C},
  \label{gH.16}
\end{equation}
which yields the corresponding gauge invariant 
quantum electrodynamics. Such a formalism does not yield a tractable 
scheme for atoms and molecules, not least because of the difficulty of giving the algebra (\ref{gH.5})-(\ref{gH.6})
a concrete realization. It is therefore customary to work instead with 
the canonical variables which have the usual canonical Poisson-brackets since in the 
original Lagrangian formulation the particle and field variables were assumed to be independent of 
each other. One then has to take account of the following equation of constraint:
\begin{equation}
  \nabla \cdot \gmat{\pi}~+~\rho~\approx 0,
  \label{gH.6a}
\end{equation}
which must be written as a `weak' equation in Dirac's terminology\cite{PAMD:52} 
if one uses the canonical variables (${\bf a}, {\gmat{\pi}}$) because (\ref{gH.6a}) interpreted as an ordinary equation is not consistent\cite{SW:13} with the canonical commutation relation  for ${\bf a}$ and 
$\gmat{\pi}$.

Now consider the general linear superposition of (\ref{gH.6a}) with a suitably smooth function $f$,
\begin{equation}
  G = \int f (\nabla \cdot \gmat{\pi}~+~\rho)\dd \tau,
  \label{gH.17}
\end{equation}
which may be used as generator of an infinitesimal canonical transformation of a dynamical 
variable $\Omega$ according to the usual rule
\begin{equation}
  \delta \Omega = \epsilon\{G,\Omega\},
  \label{gH.18}
\end{equation}
where $\epsilon$ is infinitesimal. Then we find,
\begin{alignat}{6}
&{\bm{\pi}}&\rightarrow~ &{\bm{\pi}}'~&=&~{\bm{\pi}},\\
\label{121}
&{\bf a}&\rightarrow~ &{\bf a}'~&=&~{\bf a}+\epsilon{\bf\nabla}f,\\
&{\bf x}_n&\rightarrow~ & {\bf x}_n'~&=&~{\bf x}_n,\\
\label{122} 
&{\bf p}_n&\rightarrow~ &{\bf p}_n'~&=&~{\bf p}_n-\epsilon e_n {\bf\nabla }f({\bf x}_n),\\
&H&\rightarrow~ &H'~&\equiv &~H.
\end{alignat}
Evidently $G$ is a non-trivial constant of 
the motion, and so describes an invariance of the system; it induces a 
gauge transformation in the vector potential and a corresponding change in 
the particle momenta such that the \{$\overline{\bf p}_{n}$\} are unchanged. 
It is this invariance of the physical dynamics that is referred to as \textit{gauge invariance}.

A gauge condition is any linear functional of the vector potential,
\begin{equation}
  l({\bf a}) = 0,
  \label{gH.19}
\end{equation}
which can be viewed as another constraint on the dynamical variables. If $l$ is chosen such that its Poisson-bracket with $G$, equation (\ref{gH.17}), is non-zero
\begin{equation}
  \{G(\gmat{\pi},\rho),l({\bf a})\} ~\neq~0
  \label{gH.20}
\end{equation}
it may be shown that it is possible to redefine the Poisson brackets 
of the dynamical variables (to give so-called `Dirac brackets') so that the modified brackets of the constraints
with all dynamical variables vanish identically, \textit{and} the equations 
of motion are preserved\cite{PAMD:52}. The constraints can then be taken as ordinary 
equations, and (\ref{gH.6a}) written as an ordinary equality instead of as a `weak' equality. 
In this form it simply represents Gauss's Law relating the electric field, ${\bf E}$, to the charge 
density, $\rho$,
\begin{equation}
\epsilon_{0}\nabla\cdot{\bf E}~=~\rho.
\label{GLaw}
\end{equation}
The Dirac brackets are used in exactly the same way as the usual Poisson form. After standard canonical quantization they provide the commutation relations. Thus the Hamiltonian has a fixed form, and 
every distinct gauge condition has its own set of Dirac brackets arrived at through Dirac's construction.

While there are infinitely many possible gauge conditions, Eq.(\ref{gH.19}), only two have found 
practical utility in atomic and molecular physics, the `Coulomb' or `radiation' gauge defined by
\begin{equation}
    \nabla\cdot{\bf A}({\bf x})~=~0~~~~~~~~~~\mbox{Coulomb gauge condition}
    \label{Cgauge}
\end{equation}
and the `multipolar' gauge defined by
\begin{equation}
    {\bf x}\cdot{\bf a}({\bf x})~=~0~~~~~~~~~\mbox{multipolar gauge condition}.
    \label{Mgauge}
\end{equation}
There is a long history for the use of the gauge condition (\ref{Mgauge}) and a clear connection with the PZW transformation \cite{RGW:73,RGW:74,CM:95}. Having said that, it is important to recognize that the 
PZW Hamiltonian (\S \ref{PZW}) makes no use of the vector potential. 

The `multipolar' gauge vector potential ${\bf a}_{M}$ is
defined by\cite{JDJ:02}
\begin{equation}
  {\bf a}_{M}({\bf x}',t) = {\bf A}({\bf x}',t) - \nabla_{{\bf x}'}\int_{0}^{1}\!\!\dd u~{\bf x}'\cdot{\bf A}(u{\bf x}',t)
  \label{1.1}
\end{equation}
and we use ${\bf A}$ specifically to represent the Coulomb gauge vector potential. The substitution
\begin{equation}
 {\bf z} = {\bf x}'~u, \dd{\bf z} = {\bf x}'~\dd u 
 \label{1.2}
\end{equation}
 leads to 
 \begin{equation}
    {\bf a}_{M}({\bf x}',t) = {\bf A}({\bf x}',t) - \nabla_{{\bf x}'}\int_{0}^{{\bf x}'}\!\!\!\dd{\bf z}\cdot{\bf A}({\bf z},t). 
    \label{1.3}
 \end{equation}  
This representation of the vector potential can be related to the PZW formalism in the following way\cite{RGW:03}; let
\begin{equation}
   \int_{0}^{{\bf x}'}\!\!\dd{\bf z}\cdot{\bf A}({\bf z},t) = \int~{\bf g}({\bf x}, {\bf x}')\cdot{\bf A}({\bf x})\dd{^3}{\bf x},
   \label{1.4}
\end{equation}
where
\begin{equation}
    {\bf g}({\bf x},{\bf x}') = \int_{0}^{{\bf x}'}\!\!\dd{\bf z}~\delta^3({\bf x}-{\bf z})
    \label{1.5}
\end{equation}
is a Green's function for the divergence equation,
\begin{equation}
    \nabla_{\bf x}\cdot {\bf g}({\bf x},{\bf x}') = - ~\delta^3({\bf x}-{\bf x}').
    \label{1.6}
\end{equation}
Clearly only the longitudinal component of ${\bf g}$ is well-defined by (\ref{1.6}),
\begin{equation}
    {\bf g}({\bf x},{\bf x}')^{\parallel} = \nabla_{\bf x}\frac{1}{4\pi|{\bf x}-{\bf x}'|}.
    \label{1.7}
\end{equation}

The `electric polarization field' introduced by Power and Zienau\cite{PZ:59,EAP:64} is a solution of the equation
\begin{equation}
    \nabla\cdot{\bf P}({\bf x}) = - \rho({\bf x}),
    \label{1.9}
\end{equation}
where $\rho$ is the charge density (\ref{gH.3}). It is easily seen that the polarization field 
can be written with the aid of the solution of (\ref{1.6}) in the form,
\begin{equation}
    {\bf P}({\bf x}) = \int{\bf g}({\bf x},{\bf x}')~\rho({\bf x}')\dd{^3}{\bf x}'.
    \label{1.11}
\end{equation}
Although a quantity ${\bf P}$ defined by (\ref{1.9}) is understood as 
a contribution to the `displacement field' in classical dielectric theory, here 
it is best not to make a physical interpretation since from the foregoing 
its transverse component is not determined.

The `multipolar' gauge vector potential arises when the specific choice of a 
straight path of finite length from an (arbitrary) origin to the field point ${\bf x}$, 
equation (\ref{1.2}), is made for the integration path in (\ref{1.5}), and this 
choice \textit{defines} a gauge for the vector potential. On the other hand, 
the Coulomb gauge vector potential can be defined by choosing ${\bf g}$ in 
(\ref{1.4}) as the purely longitudinal form (\ref{1.7}); an integration by 
parts in the RHS of (\ref{1.4}) shows this is equivalent to 
the conventional formulation, (\ref{Cgauge}). Equation (\ref{1.5}) implies 
that (\ref{1.4}) is path-dependent, and (\ref{1.3}) can be given a more 
general interpretation as a representation of an arbitrary vector 
potential characterized by a specified path ${\cal{P}}$. It is evident that no gauge 
condition is any more significant than any other; to suppose otherwise would be like saying a particular 
direction is physically significant in a system that is rotationally invariant. Of course 
some choice has to be made in order to have practical calculation, and some choice 
may be more convenient than others. But in the end one has to demonstrate that the 
calculation satisfies gauge invariance. We do not go into 
the details of the calculation of the commutation relations for different gauges, which are given in the 
original literature\cite{PAMD:50,PAMD:51,PAMD:52,PAMD:64}, except to note specifically 
that those for the Coulomb gauge and multipolar gauge conditions were given 
in\cite{RGW:71,RGW:73,RGW:74,BL:83,CM:95}.

Now consider the Poisson brackets (\ref{gH.5})-(\ref{gH.6}) which provide the rules 
for differentiation of a function of the phase-space variables. According to 
the first Poisson bracket in Eq. (\ref{gH.6}) we may identify the gauge 
invariant variable $\overline{\bf p}$ as the generator of 
an infinitesimal translation of the particle
\begin{equation}
  \label{gH.7}
  {\bf x} \to {\bf x} + \dd {\bf x},
\end{equation}
through an infinitesimal canonical transformation with the relation
\begin{equation}
  \label{gH.8}
  \dd{\bf x} =  \{{\bf x},\overline{\bf p}\cdot\dd{\bf x}\}.
\end{equation}

An infinitesimal translation $\dd{\bf x}$ of a general phase-space 
function $\Gamma$ is given by
\begin{equation}
  \label{gH.9}
  \Gamma({\bf x}+\dd{\bf x}) = \Gamma({\bf x})+ \{\Gamma,\overline{\bf p}\cdot\dd{\bf x}\}.
\end{equation}
If one transports $\Gamma$ around an infinitesimal rectangle with 
sides $\dd{\bf x},\dd{\bf x}'$ the result after one complete circuit is a 
change in $\Gamma$ of
\begin{equation}
  \label{circB}
  \delta \Gamma = \{\Gamma, \{\overline{p}^{r},\overline{p}^{s}\}\}\dd x^{r}\dd x'^{s}.
\end{equation}
With the aid of (\ref{gH.5}) this becomes
\begin{equation}
  \label{gH.11}
  \delta \Gamma = e\{\Gamma,{\bf B}({\bf x})\cdot \dd {\bm{\sigma}}\},
\end{equation}
where the area $\dd {\bm{\sigma}}$ is
\begin{equation}
  \label{gH.12}
  \dd {\bm{\sigma}} = \dd {\bf x} \wedge \dd{\bf x}'.
\end{equation}
A non-zero value for (\ref{gH.11}) implies that translation of 
$\Gamma$ by $\dd{\bf x}$ followed by a translation of $\dd{\bf x}'$ is not 
the same as translation first by $\dd{\bf x}'$ followed by 
$\dd{\bf x}$; basic geometry dictates that successive translations 
on curved surfaces do not commute, so we conclude that classical 
electrodynamics in Hamiltonian form has a curved phase-space 
characterized by the magnetic field.

Corresponding to the infinitesimal version (\ref{gH.11}) there is a 
finite integrated form involving the integral
\begin{equation}
  \label{gH.13}
  e\int_{\cal{S}}{\bf B}\cdot \dd {\bf S}~=~e\oint_{\cal{P}}{\bf a}\cdot \dd {\bf x}, 
\end{equation}
where the first integral is taken over a surface ${\cal{S}}$ bounded by a 
closed curve ${\cal{P}}$, and we used Stokes theorem and (\ref{gH.2}). In 
the terms of differential geometry 
the infinitesimal 1-form ${\bf a}\cdot \dd {\bf x}$ is the `connection' 
that specifies how to make infinitesimal displacements in 
the phase-space, and the magnetic field ${\bf B}$ is the associated `curvature' 
of the phase-space.

These geometrical facts about the phase-space 
of charged particles in the presence of an electromagnetic field reviewed 
above have no direct consequences in classical electrodynamics. However they 
are inherited in QED after canonical quantization, and play a fundamental 
role in the quantum theory.

The formally unitary operator
\begin{equation}
  \mathsf{U} = \exp\left(\frac{ie}{\hbar}\int_{{\cal{P}}}\!{\bm{\mathsf{a}}}\cdot \dd {\bf x}\right),
  \label{gH.15}
\end{equation}
in which the vector potential operator is integrated along some 
path ${\cal{P}}$ occurs throughout QED. It is, for example, the basis of the 
description of the magnetic field Bohm-Aharonov 
effect\cite{BA:59, CBB:07, TN:08}. Feynman in his \textit{Lectures on Physics}\cite{RF:64} uses 
the infinitesimal form of $\mathsf{U}$ (the differential 1-form 
${\bm{\mathsf{a}}}\cdot\dd {\bf x}$)
to state the fundamental quantum law for a charged particle moving in 
an electromagnetic field as a change of phase of its wavefunction. 
If ${\cal{P}}$ is taken to be a closed path, ${\mathsf{U}}$ is essentially 
the gauge-invariant Wilson loop operator (cf (\ref{gH.13}))
in both abelian (QED) and non-abelian (Yang-Mills) gauge theories (see
for example\cite{RQB:12}). In his gauge-invariant QED, Dirac\cite{PAMD:55,SM:62,RGW:71a} introduced
a quantity that is essentially the Green's function ${\bf g}$ in (\ref{1.5}), and showed that a
gauge-invariant electron field operator, $\Phi$, could be obtained from the usual
gauge dependent operator $\upphi$ by writing $\Phi = \mathsf{U}\upphi$. Finally, the specific 
choices: i) ${\bm{\mathsf{a}}}$ is the Coulomb gauge vector potential, ${\bf A}$, and 
ii) the path ${\cal{P}}$ is the straight line path of finite length from some (arbitrary) 
origin to the position of a charge $e$, identifies $\mathsf{U}$ as the 
unitary operator that effects the PZW transformation\cite{RGW:71, PZ:59, WPH:82, TDG:89, RL:83, CT:98, AS:10} of the Coulomb gauge-fixed Hamiltonian for the charge $e$. Thus the PZW transformation is 
rigorously rooted in the mainstream of modern quantum electrodynamics.

\section{The Power-Zienau-Woolley Transformation}
\label{PZW}

Taking ${\bf g}$ with the straight line path in the polarization field (\ref{1.11}), and the Coulomb gauge vector potential, ${\bm{\mathsf{A}}}$ one can form 
the PZW transformation operator (\ref{1.1U}) with\cite{RGW:71},
\begin{equation}
    \mathsf{S}= \int{\bm{\mathsf{P}}}\cdot{\bm{\mathsf{A}}}\dd\tau = \sum_{n}e_{n}{\bm{\mathsf{x}}}_{n}\cdot\int_{0}^{1}{\bm{\mathsf{A}}}(u{\bf x}_{n})\dd u,
    \label{1.12}
\end{equation}
which has a multipole expansion with leading terms that coincide with the 
form given by Power and Zienau\cite{PZ:59,AW:70,RGW:71} (see also Fiutak\cite{JF:63}). $\mathsf{S}$ 
commutes with the canonical `position' field and particle operators, so the transformation yields
\begin{alignat}{6}
&{\bm{\mathsf{A}}}&\rightarrow~ &{\bm{\mathsf{A}}}'~&=&~{\bm{\mathsf{A}}},\\
\label{126}
&{\bm{\uppi}}&\rightarrow~ &{\bm{\uppi}}'~&=&~{\bm{\uppi}}~+~{\bm{\mathsf{P}}},\\
\label{123}
&{\bm{\mathsf{x}}}_n&\rightarrow~ & {\bm{\mathsf{x}}}_n'~&=&~{\bm{\mathsf{x}}}_n,\\
\label{124} 
&{\bm{\mathsf{p}}}_n&\rightarrow~ &{\bm{\mathsf{p}}}_n'~&=&~{\bm{\mathsf{p}}}_n+{\bf\nabla}_{{\bf x}_n}\mathsf{S},\\
\label{125}
&\mathsf{H}&\rightarrow~ &\mathsf{H}'~&=&~\mathsf{H}_{\mbox{PZW}}.
\end{alignat}
Obviously it is not a gauge transformation since $\mathsf{S}$ commutes with the Coulomb gauge 
vector potential, ${\bm{\mathsf{A}}}$; since it is unitary the commutation relation 
between ${\bm{\mathsf{A}}}$ and its conjugate ${\bm{\uppi}}$ is unchanged. $\mathsf{S}$ also 
commutes with Gauss's Law (\ref{GLaw}); the form of the Hamiltonian is modified however. It may be written
\begin{align}
\label{PZWHam}
   \mathsf{H}_{\mbox{PZW}} = &\sum_{n}\frac{|{\bf p}_{n}|^2}{2m_{n}} - \int\!{\bm{\mathsf{P}}}\cdot{\bm{\mathsf{E}}}^{\perp}\dd\tau - \int\!{\bm{\mathsf{M}}}\cdot{\bm{\mathsf{B}}}\dd\tau\nonumber\\ +& \int\!\!\int\!\!{\bm{\mathsf{\cal{X}}}}:{\bm{\mathsf{B}}}{\bm{\mathsf{B}}}\dd\tau \dd\tau' + \tfrac{1}{2\epsilon_{0}}\int{\bm{\mathsf{P}}}\cdot{\bm{\mathsf{P}}}\dd\tau\nonumber\\ +&\tfrac{1}{2}\epsilon_{0}\int\bigg(|{\bm{\mathsf{E}}}^{\perp}|^2 ~+~
c^2|{\bm{\mathsf{B}}}|^2\bigg)\dd\tau
\end{align}
in the usual notation. Equation (\ref{PZWHam}) displays a remarkable feature of the transformation, namely
that, unlike the Coulomb gauge theory, there are \textit{no} explicit Coulombic energies between 
the charges. This will be discussed below. If one takes the polarization field in its general path-dependent form one has a family of unitary transformations that lead to the same form for the transformed Hamiltonian.

There are other routes\cite{EAP:78,PT:78} to this result which should be mentioned; the transformation can be carried out
at the classical level followed by standard canonical quantization. If the operator $\mathsf{S}$ 
is expressed in terms of the classical polarization field ${\bf P}$ and the Coulomb gauge vector 
potential, ${\bf A}$, it can be regarded as the generator of a classical canonical transformation of 
the (classical) Coulomb gauge Hamiltonian which results in the form (\ref{PZWHam}) expressed in classical 
variables. Alternatively, one can take the total time derivative of this classical generator and subtract
it from the Lagrangian function (\ref{gH.1}) (in an arbitrary gauge) to give a new classical Lagrangian function
\begin{equation}
\label{PZWHam}
L_{\mbox{PZW}} - \frac{\mathrm{d}}{\mathrm{d}t}\int {\bf P}\cdot{\bf a}~\dd\tau
\end{equation}
which yields the same equations of motion. The application of the general Hamiltonian method 
summarized in \S \ref{genHam} to the new Lagrangian 
leads directly to the classical form of (\ref{PZWHam})\cite{RGW:75a}; the constraint (\ref{gH.6a}) is 
replaced by the modified form
\begin{equation}
    \nabla\cdot\gmat{\pi} ~\approx ~0
\label{newcon}
\end{equation}
which has the same content as the original one, by virtue of (\ref{1.9}) and (\ref{126}). Canonical
quantization of the particle \textit{and} field variables leads directly to (\ref{PZWHam}).

In (\ref{PZWHam}) the first term is a sum of kinetic energy operators for the charges, and the last term is the usual
Hamiltonian operator for free radiation: ${\bm{\mathsf{M}}}$ is a magnetization 
density which, like ${\bm{\mathsf{P}}}$ is linear in the charge $e$, and ${\bm{\mathsf{\cal{X}}}}$ is 
a generalized diamagnetic susceptibility tensor that is proportional to $e^2$. Their particular forms 
depend on the choice made for the electric polarization operator ${\bm{\mathsf{P}}}$ through the evaluation of the gradient term in (\ref{124})\cite{RGW:71,PT:71,BPT:73,BPT:74}. 
It remains to discuss the terms involving ${\bm{\mathsf{P}}}$. If we combine (\ref{1.5}), in its generalized form for an arbitrary path ${\cal{P}}$, and (\ref{1.11}), the 
classical polarization field appears as
\begin{equation}
    {\bf P}({\bf x}) = \sum_{n}e_{n}\int_{{\cal{P}}}^{{\bf x}_{n}}\delta^3({\bf x}~-~{\bf z})\dd{\bf z}.
    \label{pathP}
\end{equation}
The transformation operator $\mathsf{S}$ then reduces to line integrals on arbitrary paths of 
the differential 1-form ${\bm{\mathsf{A}}}\cdot\dd{\bf z}$. The familiar multipolar formalism arises if 
the path ${\cal{P}}$ for each term in the sum 
is the straight line from some origin ${\bf O}$ fixed in the charge distribution to the position 
of particle $n$. For systems that are overall electrically neutral, ${\bf P}$ is independent 
of ${\bf O}$, and if required can be written\cite{RGW:71a,RGW:03}  purely in terms of the 
particle coordinates, and the paths ${\cal{P}}$. This simply expresses the fact that the 
origin ${\bf O}$ has no physical significance, and cannot appear in observables. In consequence, for molecules that have significant symmetry elements, it is always possible to choose a multipolar expansion taken about a position that naturally exploits the corresponding spatial symmetry relations. 

As a simple example consider the hydrogen atom with an electron, $-e$, at ${\bf x}_{1}$ and a proton, $+e$, 
at ${\bf x}_{2}$; direct calculation shows that the line integral
\begin{equation}
    {\bm{\mathsf{P}}}({\bf x}) = e\int_{{\bf x}_{1}}^{{\bf x}_{2}}\!\! \dd{\bf z}~\delta^3({\bf z}~-~{\bf x})
    \label{PHyd}
\end{equation}
is a solution of (\ref{Prho}) in this case\cite{RGW:71a}. No restriction to the straight line path between the two charges is implied here; the only points in space at which the polarization field is non-zero are those that lie on the chosen path. More generally, in the multi-particle neutral system the polarization field is only non-zero along the paths joining 
pairs of oppositely charged particles. The interaction term linear in 
${\bm{\mathsf{P}}}$ in (\ref{PZWHam}) then evidently describes 
an interaction between the individual charges mediated by the transverse component of 
the electric field evaluated purely on the path ${\cal{P}}$, thus between the electron and the proton
we have
\begin{equation}
    \mathsf{V}_{{\bf E}^{\perp}} ~=~ -e\int_{{\bf x}_{1}}^{{\bf x}_{2}}\!\!\dd {\bf z}\cdot {\bm{\mathsf{E}}}({\bf z})^{\perp}.
    \label{Eline}
\end{equation}
The elementary properties of line integrals show that the interaction energy (\ref{Eline}) depends on 
the path specified between the electron and proton.
If the transverse electric field is constant along the path \textit{and} the path is of finite length (\ref{Eline}) reduces to the familiar electric dipole interaction
\begin{equation}
     \mathsf{V}_{{\bf E}^{\perp}} ~\rightarrow~ -{\bm{\mathsf{d}}}\cdot{\bm{\mathsf{E}}}^{\perp}.
\end{equation}

For the quadratic term in Equation (\ref{PZWHam}) we can make an orthogonal decomposition
into parts that, respectively, have vanishing divergence (${\bf P}^{\perp})$ and vanishing Curl (${\bf P}^{\parallel})$, for then we can write
\begin{equation}
  \int{\bm{\mathsf{P}}}\cdot{\bm{\mathsf{P}}}\dd\tau~=~  \int{\bm{\mathsf{P}}}^{\parallel}\cdot{\bm{\mathsf{P}}}^{\parallel}\dd\tau~+~\int{\bm{\mathsf{P}}}^{\perp}\cdot{\bm{\mathsf{P}}}^{\perp}\dd\tau,
  \label{split}
\end{equation}
and it is usual to regard the two contributions in (\ref{split}) quite differently. Consider the 
example of the hydrogen atom again; we have similarly to (\ref{Eline})
\begin{equation}
    \int{\bm{\mathsf{P}}}\cdot{\bm{\mathsf{P}}}\dd\tau~=~ \frac{e}{2\epsilon_{0}}\int_{{\bf x}_{1}}^{{\bf x}_{2}}\!\!\dd {\bf z}\cdot{\bm{\mathsf{P}}}({\bf z}).
\end{equation}
Since $\nabla\wedge{\bm{\mathsf{P}}}$ only vanishes if ${\bm{\mathsf{P}}}$ is the gradient of a single-valued field (purely longitudinal) we see that the quadratic term in general is path-dependent, and this carries over to the many-particle case. From (\ref{1.7}) and (\ref{1.11}) we have the classical formula for the longitudinal component
\begin{equation}
    {\bf P}({\bf x})^{\parallel}~=~ \sum_{n}e_{n}\nabla_{{\bf x}_{n}}\left(\frac{1}{4\pi|{\bf x}~-~{\bf x}_{n}|}\right),
\end{equation}
which if combined with (\ref{1.12}) implies $\mathsf{S}=0$.  So, after quantization, (\ref{1.1U}) 
is the identity transformation that leaves the Coulomb gauge Hamiltonian unchanged; then the terms 
in ${\bm{\mathsf{M}}}$ and ${\bm{\mathsf{\cal{X}}}}$ reduce to the usual ${\bm{\mathsf{p}}}\cdot{\bm{\mathsf{A}}}$ and $|{\bm{\mathsf{A}}}|^2$ terms, the term linear in ${\bm{\mathsf{E}}}^{\perp}$ vanishes and\cite{AW:70,RGW:71}
\begin{equation}
   \frac{1}{2\epsilon_{0}} \int |{\bm{\mathsf{P}}}^{\parallel}|^2\dd \tau ~=~\sum_{n,m}e_{n}e_{m}\frac{1}{4\pi\epsilon_{0}|{\bm{\mathsf{x}}}_{n}~-~{\bm{\mathsf{x}}}_{m}|}
   \label{Ppar}
  \end{equation}
independently of the paths; (\ref{Ppar}) contains the usual infinite self-energy of each of the $N$ point charges which is discarded.

From the foregoing we see that the path dependence of the Hamiltonian derives from the transverse 
component of the polarization field which is \textit{arbitrary}; as far as is known there is no theory to fix its form. 
This means that the second, third, fourth and fifth terms in (\ref{PZWHam}) are also arbitrary, a fact 
that should be understood as 
an expression of the gauge symmetry of the Hamiltonian. Clearly the line integral form (\ref{pathP}) has
a transverse component determined by the path ${\cal{P}}$, and given a path one can hope to evaluate 
its contribution to (\ref{split}). It is also clear that these are formal expressions, just as 
the current-density in (\ref{gH.3}) is; they involve distributions, in the mathematical sense, expressed 
as Dirac delta functions. Multiplication of distributions in general has no meaning and so the integral over
$|{\bm{\mathsf{P}}}|^2$ cannot be definite. To illustrate this, consider the original formulation of Power and Zienau\cite{PZ:59}
\begin{equation}
    {\bm{\mathsf{P}}}({\bf x}) \approx ({\bf d}+\ldots)~\delta^3({\bf O}~-~{\bf x}),
    \label{PZdip}
\end{equation}
where ${\bf O}$ is the arbitrary origin about which the multipole expansion is made, and confine 
attention to just the electric dipole term. Then
\begin{equation}
    \int{\bm{\mathsf{P}}}\cdot{\bm{\mathsf{P}}}\dd\tau = ({\bf d}\cdot{\bf d}+\ldots)~\delta^3(0),
\end{equation}
and $\delta^3(0)$ is not defined.  If this calculation is done with the straight-line path (i.e. the multipoles summed to infinite order expressed as an integral) then various singular terms arise, but it is not clear that the full integral (\ref{split}) provides the Coulomb energy between the charges (the singular terms being supposed `renormalized' away)\cite{RQB:12,RGW:71a}. The conventional practice has always been to put (\ref{Ppar}) into the atomic/molecular Hamiltonian which, together with the free-field Hamiltonian, specify the `unperturbed' part of the problem, and to regard the second term on the RHS of (\ref{split}) as part of the (arbitrary) `perturbation'.

This singular behaviour is traceable to the \textit{point charge model} used
in the original classical formulation of electrodynamics and the associated paths \{${\cal{P}}$\} which 
are infinitely thin; the theory however is non-relativistic which means that 
photon momenta in interactions with a particle of mass $m$ must be cut-off at 
some $k_{\mbox{max}} << mc/h$, or equivalently that distances less than the Compton wavelength for the 
particle $h/mc$ must be excluded. Then all the terms in the transformed Hamiltonian are well-behaved 
(this is also true of the Coulomb gauge Hamiltonian of course), 
though arbitrary because the paths ${\cal{P}}$ are arbitrary. So how can a family of apparently arbitrary
Hamiltonians give the same results for observables? The key is that we are dealing with unitary
transformation.

The circumstances in which the Coulomb gauge and PZW Hamiltonians lead to equivalent results was 
comprehensively described in numerous publications over the years\cite{EAP:78,PT:78,WPH:77,HW:78,PT:83,AM:84,PT:85,AS:97,PT:99}. These efforts proved that the two formulations give the same results for physical observables `on the energy shell', which means identical 
results for any process that conserves energy. It is simply an aspect 
of the fact that in quantum theory, a unitary transformation of a Hamiltonian leaves 
certain quantities invariant. 
If the Hamiltonian has any discrete spectrum, then the eigenvalues are unchanged; if there 
is a continuous portion to the spectrum then it is appropriate to refer to the Hamiltonian's 
S-matrix, which is invariant. In the case of non-relativistic QED the only eigenvalue is the ground state energy, here optionally set to zero, which is invariant. Of course the S-matrix is on the 
energy-shell, so that Coulomb-gauged and PZW obtained cross-sections are the same - see for 
example explicit demonstrations for a general version of the Kramers-Heisenberg dispersion formula not restricted to the electric dipole approximation\cite{WPH:77,HW:78}. 

It is clear in these and other references cited relating to the foundations of the subject, 
that the atomic/molecular Hamiltonian and the field Hamiltonian can be taken to be the same in both the Coulomb gauge and the PZW transformed representation, but that the interaction terms are different.  On the energy shell both Hamiltonians, as well as the entire class of equivalent Hamiltonians \cite{EAP:78}, 
give identical results - and in perturbation theory, that is true to all orders\cite{RGW:00}. 
There are many instances in the literature of the equivalence of, for 
example, light scattering cross-sections, electromagnetic energy densities, and the Poynting 
vector calculated with various unitarily equivalent 
Hamiltonians\cite{WPH:77, HW:78, EAP:78, PT:78, PT:83, PT:85, PT:99, AM:84, AS:97}. 

For off-energy-shell processes (typically transient processes, within a timescale determined by energy-time uncertainty) one can obtain different
answers because unitary equivalence can no longer be guaranteed. Indeed it should be recalled that 
the original motivation of Power and Zienau\cite{PZ:59} was the observation that the lineshape 
measured in the classic Lamb shift experiments on the hydrogen atom did not agree with that 
calculated using Heitler's time-dependent perturbation theory with the 
\textit{Coulomb gauge} Hamiltonian. They found much better agreement with their 
transformed Hamiltonian using their approximate multipolar form (\ref{PZdip}) for the polarization 
field ${\bm{\mathsf{P}}}$ in (\ref{1.12}), and noted that the experiment was sufficiently accurate 
to discriminate between the two calculations. On that basis they claimed the PZW Hamiltonian gave 
a better representation of a neutral collection of charges (atom/molecule) interacting with
the quantized electromagnetic field. Of course this is not an entirely satisfactory position because an observable ought to be calculated in a \textit{gauge-invariant} fashion.

\section{Discussion}
\label{Discuss}

Our reappraisal of the fundamental theory related to the PZW Hamiltonian and associated gauge conditions in \S \ref{genHam} and \S \ref{PZW}, puts us in the position to help demystify some common misconceptions surrounding the nature of light-matter interactions within the framework of quantum electrodynamics. For example, the recent RF study\cite{RF:17} suggested that the PZW Hamiltonian is problematic because the 
Coulomb gauge (`minimal coupling') and PZW Hamiltonians predict physically different results, and 
that the Coulomb gauge Hamiltonian is to be preferred. Furthermore it was asserted that the PZW and Coulomb gauge Hamiltonians are related through a unitary transformation that does not fulfill gauge fixing constraints, and hence the former gives rise to non-physical states described as `longitudinal photon states'. It should be observed that their presentation of the 
PZW Hamiltonian is highly idiosyncratic, in that it is expressed in terms of the Coulomb gauge 
and multipolar gauge vector potentials, and the scalar potential reappears. To our knowledge 
the originators of the PZW transformation never used such a formulation; examination of the 
original literature substantiates this view\cite{PZ:59,AW:70,RGW:71}.

The reappearance of a scalar potential is certainly not in the spirit of Dirac's generalized 
Hamiltonian theory for systems with constraints; in \textit{Les transformations de jauge 
en \'{E}lectrodynamique}\cite{PAMD:52} Dirac described the mathematical basis of the relevant Hamiltonian theory and applied 
it to three cases involving the electromagnetic field: the free-field, the 
field in interaction with a massive spin-zero field (the Pauli-Weisskopf model), and with relativistic electrons. In each case Dirac arrived at the appropriate Hamiltonian and was moved to remark\cite{trans} ``\textit{Thus we may entirely neglect the dynamical variables $A^{o}$ and $B_{o}$} [$\phi$ and $\pi_{0}$ respectively in our notation]\textit{, and write the Hamiltonian $\ldots$ without modifying the equations of motion. The variables $A^{o}$ and $B_{o}$ no longer have any physical meaning}''. Dirac's insight was basically that singular Lagrangians occur when there are more dynamical variables specified
than required, and his method is simply a technique for removing the redundant variables. In the case of electrodynamics in Hamiltonian form the redundant variables are the scalar potential $\phi$ and its formal conjugate, $\pi_{0}$. When they are removed the Hamiltonian can be used in the usual way, and all the usual consequences of quantum theory apply. The discussion in \S \ref{PZW}
surely refutes the claims in RF\cite{RF:17}; we simply refer to the notion of unitary equivalence and 
point to the irrelevance of a gauge for the vector potential which does not appear in (\ref{PZWHam}). 
There is no sense in which the PZW version of electrodynamics based on the unitary transformation generated 
by $\mathsf{S}$, equation (\ref{1.12}) (or more generally, gauge theories) proves deficient.

A key feature of the PZW Hamiltonian is the absence of longitudinal components in the electric field.  As such, the quanta of the radiation field are correctly regarded as `transverse' photons.  All interactions between material particles are therefore mediated by the exchange of virtual photons with this specific transverse character\cite{AS:15,AK:15,AS:16}.  Indeed, this feature is the origin of the widespread application of the multipolar PZW development to intermolecular interactions\cite{AS:10}. However it is important to recognize that the transversality condition relates to fields that are orthogonally disposed with respect to the wave-vector.  For near-field interactions, i.e. those proximal to a source, electrodynamic coupling involves a distribution of wave-vectors with a variance in \textbf{k} expressing momentum-space uncertainty.  Transversality with respect to spatial displacement is therefore not assured.  The distinction becomes most evident in near-zone behaviour, and emerges in calculations that entail a sum over all virtual photon modes. Examples of such processes include resonance energy transfer\cite{DLA:04,AS:05,MPEL:14,JEF:14}, the van der Waals dispersion force\cite{CT:98,AS:10,AS:16}, and radiation-induced inter-particle interactions\cite{TT:80,BA:05,AS:06}. In consequence there are additional couplings that are longitudinal with respect to the displacement \textbf{x} from the source.  

This distinction between transversality with respect to \textbf{k} and with respect to \textbf{x} is crucial. It is confusing to suggest  that `longitudinal polarization' is non-physical, because it fails to make clear with which direction the sense of the term is meant. It does appear that RF confuse `longitudinal' with respect to position ${\bf x}$, (at the heart of their representation of the Poincar\'{e} or multipolar gauge), with its meaning with respect to wave-vector, ${\bf k}$.  As observed above, the multipolar gauge vector potential has a non-zero longitudinal part because of its definition involving the gradient of the PZW transformation operator. In fact such contributions are annihilated by the physical magnetic field occurring in (\ref{PZWHam}) and of course only the transverse component of the electric field is required in the interaction and free-field terms.

Another apparent misconception arises in the view that only if one goes to the electric dipole approximation does the interaction operator reduce to a term linear in the field variables and proportional to the electric charge $e$.\cite{RF:17} As has been seen, expansions of the second and third terms of equation (47) yield the electric and magnetic multipole series, from which the prominence of the electric dipole coupling term readily follows on making the long wavelength approximation. The third interaction term in the PZW Hamiltonian is proportional to the square of the electronic charge and depends bilinearly on the magnetic field, giving rise to the diamagnetic coupling term\cite{PZ:59,RGW:71,CT:98,AS:10,AS:15,KAF:15}. Only the interaction terms that are linear in the Maxwell fields are employed in first-order processes such as one-photon absorption. Describing higher-order effects proportional to nonlinear powers of the electromagnetic field, as occur in multiphoton phenomena, poses no problem as they are treated by successive application of field operators within a desired order of multipole moment approximation. The same is true when performing calculations with the Coulomb gauge Hamiltonian. For single-photon absorption and emission (stimulated and spontaneous) of light, there is no contribution to the matrix element from the interaction term that is proportional to the square of the vector potential, with only the coupling term that is dependent upon the momentum needing to be retained. Both interaction terms, however, must be kept in calculations of processes that involve annihilation or creation of two or more photons, or the absorption of one photon and the emission of another, as occurs in linear light scattering. Hence both coupling terms dependent on the vector potential must be used when evaluating higher-order processes, irrespective of whether or not the field is spatially uniform. This is essential in guaranteeing gauge invariance.

In this context it is noteworthy that a special case of the PZW transformation has recently been applied to the Coulomb gauge Hamiltonian so as to eliminate the vector potential squared term and the static dipolar coupling\cite{VGD:14,VGD:15}. This is done in order to study the onset of ultrastrong coupling of radiation with atoms and molecules in confined geometries. It is worth pointing out that a general theory of quantum electrodynamics in a cavity has been available for some time\cite{PT:82}; a canonical transformation of the Coulomb gauge Hamiltonian with the generator equation (41) yields a PZW Hamiltonian with polarisation, magnetisation and diamagnetisation distributions coupled to Maxwell fields as in the free space case. Interestingly, performing a PZW transformation results in the complete elimination of the Coulomb interaction and that there are no contributions from image charges. It has also been illustrated, in work on the linear electro-optic effect, how the effect of an external static field can be accommodated by formulating theory for an entire system including a notional dipolar source.  With the usual device of summing over virtual photon couplings, based on the PZW Hamiltonian the emerging result duly represents all of the radiation interactions in terms of real photons, whilst the field dependence exhibits the correct classical form.\cite{LCD:00,DLA:04}
 
\section{Outlook}
\label{out}
The increasingly wide sphere of applications in which the PZW Hamiltonian is now deployed has provided an opportunity to reevaluate, from a position of current understanding, the origins and development of the underlying theory.  The analysis of its position within a generic framework of unitary transformations establishes its reliability and its relation to other gauge theories. In addressing optical interactions the PZW form offers significant calculational advantages and insights: the couplings between the optical fields and charges are defined in terms of physically intuitive electric and magnetic fields.This in turn enables a clear and direct linkage to the multipole moments and optical response tensors involved in optical phenomena, in a cast that elucidates their connections with molecular symmetry. This reassertion of validity provides a timely and necessary corrective, consolidating the ground to advance future applications.

As we approach the second quarter of the 21st century, there are a number of directions that research in quantum optical interactions can be seen to be heading. Some are centred around a better understanding of fundamental aspects of light-matter interactions, such as processes related to cavity QED physics \cite{LG:15} and quantum relaxation and decoherence \cite{SKA:14}. Others are focused on intermolecular interactions such as absorption, excitation energy transfer, Casimir - van der Waals forces, and nonlinear optics. Recent work has examined the effect of including one or more additional particles in modifying these interactions \cite{GJD:02,BKMM:09,AK:12,AK:13,MPEL:14,AS:14,WPM:15,DLA:13,JSF:13,MMC:14,JML:14} by extending the theory to account for three- and many-body couplings, thereby going beyond the pairwise additive approximation and exploring connections between microscopic and macroscopic manifestations of these processes.

Especially interesting is the application of results derived and insight gained from the solution of the problem of resonance energy transfer between a pair of atoms or molecules to the dispersion interaction between one ground and one excited state species \cite{GJ:97,SK:15,PB:15,MR:15,MD:16,PB:16}, in which fundamental issues surrounding the nature of a potential versus a rate, the functional form of the resonant term, and whether transfer of excitation energy is reversible or not, have all been addressed. In both the migration of energy and the van der Waals dispersion interaction, the starting point in the calculation is the PZW Hamiltonian. Expression of the interaction Hamiltonian in terms of multipole moments coupled to Maxwell fields is also advantageous when evaluating phenomena involving chiral species \cite{PB:17}, often demanding relaxation of the electric dipole approximation and the inclusion of effects due to magnetic dipole and electric quadrupole coupling. A more complete understanding of these processes will lead to design principles that are expected to give rise to the development of new nano-materials for light-harvesting technologies and photovoltaics. Quantum optical theories offer a complete description of radiation-induced inter-particle forces \cite{TT:80,BA:05,AS:06} thereby allowing realization of optical binding,\cite{DZ:10} a technique currently in its infancy - although first predicted using PZW theory\cite{TT:80} - which promises new applications in materials, biological and medical research. Optical chirality, \cite{TC:10} twisted light and exotic forms of electromagnetic radiation\cite{LA:99,MJP:11,AMY:11,KYB:11} offer the possibility of new advances in cutting-edge imaging and information processing technologies. Recently there has been interest in the use of quantum light as a way of probing electrodynamical couplings in chemical systems (such as in FRET systems)\cite{KB:14, ZZ:17}. The development of such theory, along with its experimental realization, could open new windows into understanding dynamical processes occurring in condensed phase quantum systems. 

All of these areas of research are highly active across a variety of sub-disciplines within the chemical physics and optics communities. There is now a strong trend towards synergistic approaches to scientific and technological development, in which the strategic design of experimental work and the implementation of computational modelling are both underpinned by the application of fundamental theory. To ensure continued success and innovation in the spheres of quantum optics, photonics and molecular physics, it is more important than ever to deploy theory that firmly stands on solid foundations.  The use of quantum electrodynamics based on the PZW Hamiltonian guarantees accuracy and success.

\end{document}